\documentclass[11pt]{article}
\pdfoutput=1
\usepackage{amsmath}
\usepackage{amssymb}
\usepackage{fancyhdr}
\usepackage{bbold}
\usepackage{appendix}
\usepackage[sort&compress,square,comma,numbers]{natbib}

\usepackage[
	colorlinks=true,
	citecolor=black,
	linkcolor=black,
	urlcolor=blue,
	hypertexnames=false]{hyperref}
	
\newcommand{\arXiv}[2]{\href{http://arxiv.org/pdf/#1}{{\tt #2/#1}}}
\newcommand{\arXivold}[1]{\href{http://arxiv.org/pdf/#1}{{\tt #1}}}

\let\oldsqrt\sqrt

\newcommand{\beq}{\begin{eqnarray}}
\newcommand{\eeq}{\end{eqnarray}}

\def\sqrt{\mathpalette\DHLhksqrt}
\def\DHLhksqrt#1#2{%
\setbox0=\hbox{$#1\oldsqrt{#2\,}$}\dimen0=\ht0
\advance\dimen0-0.2\ht0
\setbox2=\hbox{\vrule height\ht0 depth -\dimen0}%
{\box0\lower0.4pt\box2}}
\allowdisplaybreaks[1]

\renewcommand{\part}[1]{\vspace{.10in}{\bf (#1)}}
\newcommand{\Kahler}{K\"ah\-ler~}

\begin{document}

\begin{titlepage}

\vskip.5cm \begin{center} {\huge \bf Marginal Breaking of  \\
\vskip 8pt
 Conformal SUSY QCD}\\ \vskip
8pt

\vskip.1cm \end{center} \vskip0.2cm

\begin{center} {\bf {Kevin F. Cleary} {\rm
and} {John Terning}} \end{center} \vskip 8pt

\begin{center} {\it Department of Physics, University of California, Davis, CA
95616} \\

\vspace*{0.3cm} {\tt cleary@ms.physics.ucdavis.edu, terning@physics.ucdavis.edu} \end{center}

\vglue 0.3truecm

\begin{abstract}
\vskip 3pt
\noindent 
We provide an example of a 4D theory that exhibits the Contino-Pomarol-Rattazzi mechanism,  where breaking conformal symmetry by an almost marginal operator leads to a light pseudo-Goldstone boson, the dilaton, and a parametrically suppressed contribution to vacuum energy. We consider SUSY QCD at the edge of the conformal window and break conformal symmetry by weakly gauging a subgroup of the flavor symmetry.  Using Seiberg duality we show that for a range of parameters the singlet meson in the dual  theory reaches the unitarity bound, however, this theory does not have a stable vacuum. We stabilize the vacuum with soft breaking terms, compute the mass of the dilaton,   and determine the range of parameters where the leading contribution to the dilaton mass  is from the almost marginal coupling. 
\end{abstract}

\end{titlepage}

\section{Introduction}
A dilaton is a (pseudo-)Goldstone mode associated with the spontaneous breaking of scale invariance. The dilaton can be massless if the spontaneously generated scale  is a flat direction \cite{Fubini:1976jm},  or it can be a pseudo-Goldstone boson if the scale is an approximate flat direction and there is some small explicit conformal breaking by a nearly marginal operator, as in the Contino-Pomarol-Rattazzi mechanism \cite{CPR,Bellazzini:2013fga}. The coupling of the nearly marginal operator must remain small throughout its slow renormalization group (RG) running so as to not break scale invariance too badly. This is why QCD-like theories do not have a light dilaton \cite{Bellazzini:2013fga,Holdom:1987yu}, since the gauge couplings becomes large and rapidly running near the scale of the condensate $\Lambda_{QCD}$.\

A theory consisting of particles with spins $\le 1$ is conformal if there are no dimensionful couplings that appear in the theory (meaning that all operators are marginal) and if the couplings do not run. To break conformal symmetry softly we can introduce an almost marginal operator $\Delta\left(\mathcal{O}\right) = 4-\epsilon$, with $\epsilon \ll 1$, which introduces an explicit mass scale. In a dual AdS$_5$ description, where a Goldberger-Wise field \cite{Goldberger:1999uk} plays the role of the bulk field dual to the almost marginal coupling, one can explicitly perform the calculation \cite{Bellazzini:2013fga} and see that the dilaton/radion is light at the minimum of its potential. Here we want to study this scenario using a purely 4D description. Seiberg duality \cite{Seiberg:1994pq} and analytic continuation in superspace \cite{ArkaniHamed:1998kj,ArkaniHamed:1998wc,Luty:1999qc} will allow us enough theoretical control to see that weakly gauging a global symmetry of a particular conformal field theory (CFT) results in a light dilaton.

In SUSY gauge theories the  $R$-current can mix with any global $U(1)$, but if the theory is also a CFT then there is a unique combination that enters the superconformal algebra and fixes the dimension of an operator $\mathcal{O}$ in terms of  its superconformal $R$-charge:
\begin{equation} \Delta\left(\mathcal{O}\right) = \frac{3}{2}R_{sc}\left(\mathcal{O}\right) ~. \end{equation}
This is a useful tool in obtaining the edges of the ``conformal window". The unique, anomaly free $R$-charge for quarks in supersymmetric QCD (SQCD) is given by $R\left(Q\right)=1-N/F$, this tells us that if we want free quarks, meaning quarks which have $\Delta\left(Q\right) = 1$ this requires that $F=3N$. Using Seiberg duality, to write down a theory of composite mesons we can similarly reason that for a dual theory of free mesons $F=3N/2$; this gives us the bounds on the conformal window, $3N \geq F \geq 3N/2$. Identifying the mesons in the magnetic theory with bilinears of quarks in the electric theory we see that the $R$-charge of the meson receives corrections associated with the anomalous dimension of the quarks at the infrared fixed point. 
\begin{equation} R(M) = \frac{2}{3} \left(2 + \gamma_*\right) \end{equation}

When the meson of a conformal SQCD theory gets a VEV in the moduli space of vacua, it plays the role of a massless dilaton.
The natural question this raises is: can we introduce additional, almost marginal, interactions that change the anomalous dimensions of the quarks, such that we can find a unique vacuum, as in the Contino-Pomarol-Rattazzi mechanism, where  the meson is a light dilaton? We will see that by gauging a subgroup of the flavor symmetry in SQCD this can indeed be accomplished. 

In this paper we examine the almost marginal breaking of conformal symmetry by weakly gauging (with strength $g'$) a flavor subgroup of SQCD. We show that while this does cause singlet mesons in the magnetic theory to approach the unitarity bound, $\Delta \ge 1$, this theory does not have a stable vacuum. We then introduce soft SUSY breaking  in order to stabilize the vacuum and via the method of analytic continuation in superspace compute the dilaton mass. We find that there is a soft mass induced at first order in $g'^2$ due to the one-loop running of the SUSY breaking mass which is the leading effect for a range of SUSY breaking masses.\

This paper is organized as follows. First, in section 2, we discuss the weakly gauged version of SQCD and its dual; we then examine the RG flow of dual couplings at the bottom edge of the conformal window, and find how the fixed point values for the Yukawa couplings depend on $g'$. In section 3 we review the method of analytic continuation in superspace, and use it to compute the vacuum state and the dilaton mass, which arises at $\mathcal{O}\left(g'^2\right)$. In an appendix we list the renormalization group equations (RGEs) for weakly gauged SQCD as well as the fixed point solutions at $F=3N/2$.
\section{Weakly Gauged SQCD and its dual}
\subsection{Review of SQCD}
First we will briefly review the dual description of SQCD for $F> N+1$. The matter content for SQCD with $N$ colors is given by the table below:
\beq
\begin{tabular}{ c|c|cccc } 
 & $SU\left(N\right)$  & $SU\left(F\right)$ & $SU\left(F\right)$ & $U\left(1\right)_B$ & $U\left(1\right)_R$ \\ 
  \hline
$ Q $& $\Box$ & $\Box$ & 1 & 1 & $\frac{F-N}{F} \vphantom{\sqrt{\frac{N}{F-N}}}$\\ 
$ \overline{Q} $& $\overline{\Box}$ &1 & $\overline{\Box}$ & -1&$\frac{F-N}{F} \vphantom{\sqrt{\frac{N}{F-N}}}$\\ 
\end{tabular}~.
\eeq
The $R$ charge assignments are fixed by requiring that the anomaly associated with the insertion of one $R$ current and two gauge currents vanishes. This theory is asymptotically free in the UV for $F<3N$, and it becomes strongly coupled at an intrinsic scale $\Lambda$. A low-energy effective description is available to us through the use of Seiberg duality \cite{Seiberg:1994pq}. The matter content of the IR theory is:
\beq
\begin{tabular}{ c|c|cccc } 
 & $SU\left(F-N\right)$  & $SU\left(F\right)$ & $SU\left(F\right)$ & $U\left(1\right)_B$ & $U\left(1\right)_R$ \\ 
  \hline
$ q $& $\Box$ & $\overline{\Box}$ & 1 & $\frac{N}{F-N}$ & $\frac{N}{F}  \vphantom{\sqrt{\frac{N}{F-N}}}$\\ 
$ \overline{q} $& $\overline{\Box}$ & 1 & $\overline{\Box}$ & $-\frac{N}{F-N}$&$\frac{N}{F} \vphantom{\sqrt{\frac{N}{F-N}}}$\\ 
$M$ & 1 & $\Box$ & $\overline{\Box}$ & 0 & $2\frac{F-N}{F}\vphantom{\sqrt{\frac{N}{F-N}}}$\\
\end{tabular}~.
\eeq
This theory contains dual, or ``magnetic", quarks charged under the $SU\left(F-N\right)$ as well as meson fields identified with the quark bilinear $M^i_j = \overline{Q}^i_n Q^n_j$. This theory admits a unique superpotential 
\beq
W=\lambda \,\overline{q} Mq~,
\label{dualsup}
\eeq
 where $\lambda$ is the Yukawa coupling between dual quarks and mesons. For $3N \geq F \geq 3N/2$ the couplings approaches IR fixed points. For sufficiently small $F$ this theory is weakly coupled in the IR, and becomes strongly coupled  at higher energies near the intrinsic scale $\Lambda$. 

\subsection{Weakly Gauged SQCD}
We now want to gauge a vector-like $SU\left(N'\right)$ subgroup of the $SU\left(F\right)\times SU\left(F\right)$ flavor symmetry. Gauging a subgroup of the flavor symmetry will split the quarks into $Q$ and $Q'$.
The charge assignments are given in the table below: 
\beq
\begin{tabular}{ c|cc|ccc } 

 & $SU\left(N\right)$ & $SU\left(N'\right)$ & $SU\left(F-N'\right)$ & $SU\left(F-N'\right)$ &$U\left(1\right)_R$ \\ 
  \hline
$ Q $& $\Box$ & 1 & $\Box$ & 1 &$1+\frac{N'^2-N^2}{N\left(F-N'\right)}  \vphantom{\sqrt{\frac{N'^2-N^2}{N\left(F-N'\right)}}}$ \\ 
$ \overline{Q} $& 1 & $\overline{\Box}$ & 1 & $\overline{\Box}$ &$1+\frac{N'^2-N^2}{N\left(F-N'\right)}  \vphantom{\sqrt{\frac{N'^2-N^2}{N\left(F-N'\right)}}}$\\ 
$ Q' $& $\Box$ & $\Box$ & 1 & 1 & $\frac{N-N'}{N}  \vphantom{\sqrt{\frac{N'^2-N^2}{N\left(F-N'\right)}}}$ \\ 
$ \overline{Q}' $& $\overline{\Box}$ & $\overline{\Box}$ & 1 & 1 &$\frac{N-N'}{N}   \vphantom{\sqrt{\frac{N'^2-N^2}{N\left(F-N'\right)}}}$ \\ 

\end{tabular}~.
\eeq
The $U\left(1\right)_R$ charges are assigned to make both gauge anomalies  of the $R$-current vanish. There are two additional $U\left(1\right)$ anomaly free charges. One is just a baryon number under which all quarks have positive charge and all anti-quarks have a negative charge. The other is a subgroup of the $SU(F)\times SU(F)$ flavor symmetry  that is left-over after gauging the vector-like $SU\left(N'\right)$ subgroup. The additional $U(1)$'s  have been omitted from the table for the sake of brevity. We now want to write down a weakly coupled dual description of this theory, using Seiberg duality \cite{Seiberg:1994pq}, which will have five kinds of mesons made out of all possible combinations of quarks and anti-quarks:
\beq
M^i_j &=& \overline{Q}^i Q_j~, \quad M^{\prime\, i}_j = \overline{Q}^{\prime i} Q^{\prime}_j~, \\
M^\prime_\Box &=& \overline{Q}^i Q^\prime_j\ ~,\quad M^\prime_{\overline{\Box}}=\overline{Q}^{\prime i} Q_j~\\
M^\prime_A &=& \overline{Q}^{\prime i} T^a Q^\prime_j~.
\eeq
where $T^a$ is a generator of $SU\left(N'\right)$.
The weakly coupled dual description of this theory is:
\beq
\begin{tabular}{ c|cc|ccc } 

 & $SU\left(F-N\right)$ & $SU\left(N'\right)$ & $SU\left(F-N'\right)$ & $SU\left(F-N'\right)$ &$U\left(1\right)_R$ \\ 
  \hline
$ q $& $\Box$ & 1 & $\overline{\Box}$ & 1 &$\frac{N'^2-N^2}{N\left(F-N'\right)}$ \\ 
$ \overline{q} $& $\overline{\Box}$ & 1 & 1 & $\Box$ &$\frac{N'^2-N^2}{N\left(F-N'\right)}$\\ 
$ q' $& $\Box$ & $\Box$ & 1 & 1 & $\frac{N'}{N}$ \\ 
$ \overline{q}' $& $\overline{\Box}$ & $\overline{\Box}$ & 1 & 1 &$\frac{N'}{N}$ \\ 
$ M $& 1 & 1 & $\Box$ & $\overline{\Box}$ &$2\left(1+\frac{N'^2-N^2}{N\left(F-N'\right)}\right)$ \\ 
$ M' $& 1 & 1 & 1 & 1 &$2\frac{N-N'}{N}$\\ 
$ M'_\Box $& 1 &$\Box $& $\Box$ & 1 &$2-\frac{N'}{N}+\frac{N'^2-N^2}{N\left(F-N'\right)}$ \\ 
$ M'_{\overline{\Box}} $& 1& $\overline{\Box}$ & 1 & $\overline{\Box}$ &$2-\frac{N'}{N}+\frac{N'^2-N^2}{N\left(F-N'\right)}$\\ 
$ M'_A $& 1 & Adj & 1 & 1 &$2\frac{N-N'}{N}$ \\ 

\end{tabular}~.
\eeq
This dual theory admits a unique superpotential:
\begin{equation} W = \lambda_1 q\overline{q}M + \lambda_2 q\overline{q}'M'_{\Box} + \lambda_3 q'\overline{q}M'_{\overline{\Box}} +\lambda_4 q'\overline{q}'M' + \lambda_5 \left(q'T^a\overline{q}'\right) M'^a_A~.
\end{equation}
 In the limit of $g'\rightarrow 0$, we should find that all the Yukawa couplings are equal as in (\ref{dualsup}).

\subsection{RG Analysis}
\label{sec:RG}

We want to examine the space of couplings $\left(g,\lambda_i\right)$ with $g'$ kept perturbative, meaning that we assume\footnote{We could also add spectators that only have $SU(N')$ gauge couplings, if needed to adjust the $\beta$ function.} its strong couling scale $\Lambda^\prime$ is much smaller that any other scale we are interested in. With the number of flavors, $F$, close to $3N/2$ the IR couplings $\left(g,\lambda_i\right)$ are also perturbative. We derive the RGEs, following ref. \cite{deGouvea:1998ft}.  The superpotential is not renormalized, so we only need to compute wavefunction renormalization. If we are exactly at the fixed fixed point then the coupling will, of course, not run. For example, this implies
\begin{equation} 
\lambda_1\left(\mu\right) = \lambda_1\left(\mu_0\right) Z_q^{-\frac{1}{2}} Z_{\overline{q}}^{-\frac{1}{2}} Z_m^{-\frac{1}{2}} = \lambda_{1*}\end{equation}
Here the wavefunction renormalization factors are given by the following expressions:
\beq Z_q &=& 1+\left(\frac{g^2}{8\pi^2} 2C_2\left(\Box\right) - \frac{\lambda_1^2}{8\pi^2} \left(F-N'\right) - \frac{\lambda_2^2}{8\pi^2} N'\right)\log \frac{\mu}{\mu_0}~,\\
Z_{\overline{q}} &=& 1+\left(\frac{g^2}{8\pi^2} 2C_2\left(\Box\right) - \frac{\lambda_1^2}{8\pi^2} \left(F-N'\right) - \frac{\lambda_3^2}{8\pi^2} N'\right)\log \frac{\mu}{\mu_0}~,\\
Z_m &=& 1-\frac{\lambda_1^2}{8\pi^2}\left(F-N\right) \log \frac{\mu}{\mu_0}~.
\eeq
These wavefunction renormalization factors lead to the RGE\footnote{The complete set of RGEs are in Appendix A.}:
\beq
\frac{d}{d \ln \mu} \lambda_1^2 = \frac{\lambda_1^2}{8\pi^2} \left(2 \lambda_1^2 \left(F-N'\right) + \lambda_2^2 N' + \lambda_3^2 N' + \lambda_1^2  \left(F-N\right) -2 g^2 \, 2C_2\left(\Box\right) \right)~.
\eeq

It is useful, as a sanity check, to take the $g'\rightarrow 0$ limit where all of the Yukawa couplings must be equal, and the RGE must be independent of $N'$. Taking this limit we see that the RGE is in fact independent of $N'$ and agrees with the known result \cite{deGouvea:1998ft}:
\begin{equation} \frac{d}{d \ln \mu} \lambda^2= \frac{\lambda^2}{8\pi^2} \left(2\lambda^2F+\lambda^2\left(F-N\right) -2 g^2\,2C_2\left(\Box\right)\right)~.\end{equation}

Having obtained the RGEs, we can, for choices of $\left(N,F,N'\right)$, compute the value of the fixed points as a function of $g'$. We find generically that for $F\ge 3N/2$ the only Yukawa couplings which vanish for a given value of $g'$ are those associated with the singlet mesons, $M$ and $M'$. Thus these mesons can become free fields,  or in other words they can reach their unitarity bound. The singlet meson made out of the $Q'$s, $M'$, always hits the unitarity bound first; first, in this context, meaning for smaller values of $g'$. The solutions for the fixed points are of the form:
\begin{equation}  
\lambda^{2}_{i*} =  \lambda^{2}_*(g'=0) + a_i \,g'^2
\end{equation}
Where  $\lambda_*(g'=0) $ is the fixed point value of the Yukawa coupling  in the absence of the coupling $g'$. The presence of $g'$ splits the fixed point values of the Yukawa couplings, and the fixed point values vary with $g'$. The meson with a Yukawa coupling with the largest negative $a_i$ will be the first to hit the unitarity bound as $g'$ increases. We are assuming that $g'$ is asymptotically free, so that (slowly) increasing $g'$ is associated with RG running to lower energy scales. 

\subsection{Vacua}
We now want to examine the space of vacua of this theory. We can introduce a VEV for the scalar component of $M'$, which we will ultimately identify with the dilaton, and investigate the theory below the scale of the  $M'$ VEV. $F$-flatness conditions have no nontrivial solutions, and a mass is generated for the dual quarks, $q'$ and $\overline{q}'$. At scales below the VEV of $M'$ the dual quarks can be integrated out of the theory. We than have  two gauge groups, $SU\left(F-N\right)$ with $F-N'$ flavors, and $SU\left(N'\right)$ with 0 flavors. So we see that the $SU\left(N'\right)$ sector is pure SUSY Yang Mills and undergoes gaugino condensation. From matching the scale $\Lambda'$ above and below the VEV of $M'$ we see that  a term is generated in the superpotential proportional to $\ln M'$:
\begin{equation} \int d^2 \theta \ln M' W'^\alpha W'_\alpha +h.c. 
\end{equation}
The corresponding scalar potential associated with this term has a minimum at $M' \rightarrow \infty$. This signals the breakdown of our effective theory since we cannot trust the IR dual at large VEVs. In order to stabilize the vacuum at small VEVs we will have to include a small soft SUSY breaking mass.
One may have expected this as recent work has found that theories with light dilatons also contain a small cosmological constant \cite{Bellazzini:2013fga}, and there is obviously no way to generate a cosmological constant in the context of unbroken SUSY. Equivalently, with unbroken SUSY a dilaton must be massless. In the next section we will include soft SUSY breaking, which we will see can stabilize the vacuum at small VEVs.

\section{Analytic Continuation in Superspace}

We next want to include soft SUSY breaking in the dual theory via the method of analytic continuation in superspace \cite{ArkaniHamed:1998kj,ArkaniHamed:1998wc,Luty:1999qc} in the presence of parametrically small SUSY breaking mass terms. To do this we introduce non-vanishing $F$ and $D$-term SUSY breaking spurions, and compute soft masses in this SUSY breaking background.
We first briefly review this formalism in SQCD, closely following the discussion in \cite{Csaki:2012fh}. The action for SQCD in the UV is given by:
\begin{equation} \mathcal{L} = \int d^4 \theta \, Q^\dagger \mathcal{Z}e^V Q + \int d^2\theta \,U \,W^\alpha W_\alpha~.
\end{equation}
We can introduce SUSY breaking spurions given by scalar and gaugino masses in the following way:
\beq  
\label{Zspurion}
\mathcal{Z} &=& Z\left(1-\theta^2\bar{\theta}^2 m_{UV}^2\right)~,\\
U &=& \frac{1}{2g^2} -i\frac{\theta_{YM}}{16\pi^2} + \theta^2 \frac{m_{\lambda}}{g^2}~.
\eeq
The holomorphic scale associated with this theory is related to $U$ by
\begin{equation} \Lambda_h = \mu \,e^{-\frac{16\pi^2 }{b}U\left(\mu\right)}~.\end{equation}
We will be interested in the case where $m_{UV}$, $m_{\lambda} \ll \Lambda$. 

Now consider the action of anomalous rescalings on this theory. For the action to be invariant under transformations $Q\rightarrow e^AQ$ the wavefunction renormalization factor must scale like a real vector superfield:
\begin{equation} \mathcal{Z} \rightarrow e^{-\left(A+A^\dagger\right)}\mathcal{Z}~. \end{equation}
The holomorphic scale will also have a nontrivial transformation due to the axial anomaly:
\begin{equation} \Lambda_h \rightarrow e^{\frac{2F}{b}A} \Lambda_h~. \end{equation}
It is also convenient to introduce a redundant scale which is axially invariant:  
\begin{equation} \Lambda^2 =\Lambda_h^\dagger \mathcal{Z}^{\frac{2F}{b}} \Lambda_h~. \end{equation}

Mesons in the dual description are identified with quark bilinears in the electric theory. This implies that the mesons in the IR theory transform under axial transformations as $M\rightarrow e^{2A} M$. Imposing SUSY and axial invariance of the low energy action the \Kahler term for the meson is given by:
\begin{equation} 
\mathcal{L}_K = \int d^4\theta \,\frac{M^\dagger \mathcal{Z}^2 M}{\Lambda^2}~.
\end{equation}
Taylor expanding in superspace this gives the $\mu \rightarrow 0$ value of the meson mass. There are nontrivial corrections at $\mu \neq 0$ that we will discuss in the next section. We find the mass of the meson given in terms of $m_{UV}$ is given by:
\begin{equation} m^2_{M} = 2\left(\frac{3N-2F}{3N-F}\right) m_{UV}^2 
\label{simplemass}
\end{equation}
We see at the bottom edge of the conformal window ($F=3N/2$) there is no soft mass associated with the meson. We now want to apply this type of analysis to weakly gauged SQCD, including the effects of RG Evolution.

\subsection{RG Evolution of Soft Masses}

We will need to evaluate soft masses at an arbitrary RG scale. In the electric description of softly broken SQCD defined at a UV scale $\mu \rightarrow \infty$, SQCD is a free theory. As we lower the RG scale, $\mu$, we introduce corrections to the soft masses due to the presence of gauge interactions. In the conformal window the corrections are fixed by anomalous dimensions.
Seiberg duality allows us to write down a low energy effective description of our theory below the strong coupling scale $\Lambda$, at which the theories have to match. The mapping of soft masses between UV and IR, without including anomalous dimensions given in Eq. (\ref{simplemass}), is correct  when  the UV theory is asymptotically free in the deep UV, and its dual, is IR free.

In our weakly gauged SQCD example, however, the IR theory flows to a non-trivial fixed point, and in addition we are interested in renormalized masses at physical thresholds rather than at $\mu=0$. Corrections to soft masses at arbitrary scales were computed in \cite{Luty:1999qc} using the method of the superconformal compensator. The method of \cite{Luty:1999qc} is to couple a SUSY gauge theory to a SUSY breaking supergravity background with a gauged $R$ symmetry via the superconformal compensator field $\phi$. There are  two $U\left(1\right)$ symmetries in this approach. In supergravity there is the gauged $R$ symmetry, with a gauge field included in the vector superfield $V_R$. The current of this gauge symmetry is the lowest member of the stress tensor supermultiplet. All matter fields have vanishing $R$ charge under this symmetry. There is another global $R$ charge under which each matter field has charge $R$ and $\phi$ has charge $R(\phi) =-\frac{2}{3}$. The VEV of $\phi$ breaks these two $U\left(1\right)$s down to the diagonal $U\left(1\right)$ and it is this subgroup which corresponds to the usual anomaly free $R$ charge. The action, which is fixed by $R$-invariance is given by:
\begin{equation} \mathcal{L} = \int d^4 \theta \left(\phi^\dagger e^{-\frac{2}{3}V_R}\phi\right) \left(Q^\dagger \mathcal{Z}e^V e^{V_R R} Q\right) + \int d^2\theta UW^\alpha W_\alpha~.
\end{equation}

The SUSY breaking supergravity background associated with the vector superfield $V_R$ and the superconformal compensator $\phi$ are given by:
\begin{equation} V_R = \theta^2 \bar{\theta}^2D_R, \quad \phi = 1 + \theta^2 F_{\phi}~.
\end{equation}
Regulating the theory necessarily introduces a scale, and this will introduce a nontrivial dependence on $\phi$ through loops. It was pointed out in  \cite{Luty:1999qc}  that the additional dependence on $\phi$ can be captured by shifting the RG scale in the following way:
\begin{equation}\mu^2 \rightarrow  \hat{\mu}^2=\frac{\mu^2}{\phi^\dagger e^{-\frac{2}{3}V_R} \phi}~.
\label{rescaling}
\end{equation}
This identification feeds the SUSY breaking background information into the RG evolution. Identifying the soft masses of these fields with the nonvanishing theta components of $\mathcal{Z}$, as in Eq. (\ref{Zspurion}), allows us to extract the SUSY breaking information by computing $\mathcal{Z}\left(\hat{\mu}\right)$. 

In general $\mathcal{Z}$ can be written as the exponential of the integral of the anomalous dimension $\gamma$. The rescaling (\ref{rescaling}) introduces a change in the measure, $d \ln \mu$, of this integral and the SUSY breaking information is fed in through this change of the measure. The full analysis \cite{Luty:1999qc} yields:
\begin{equation} m^2\left(\mu \right) = -\ln \mathcal{Z}\left(\hat{\mu}\right)\mid_{\theta^2\overline{\theta}^2} = \left(\frac{2}{3} - R - \frac{1}{3} \gamma \right)D_R -\frac{1}{4}\frac{d\gamma}{d\ln \mu} \mid F_{\phi}\mid^2~.
\label{luty}
\end{equation}

Where $\gamma = -\frac{d \ln \mathcal{Z}}{d \ln \mu}$. We see that this agrees with results from SQCD in the conformal window where $R=\frac{2}{3}\left( 2 + \gamma_*\right)$ and the fact that effects due to gauge interactions contribute negatively to the $R$-charge of the meson. (This is what allows the dimension of meson in the magnetic description of SQCD to reach the unitarity bound and for the meson to become a free field.) Taking $\gamma \rightarrow 0$ to relate the IR and UV masses, as in (\ref{simplemass}), we see that we can identify the SUSY breaking parameters of the two approaches by 
\beq
D_R = \frac{2F}{b}m_{UV}^2~.
\eeq

\subsection{Perturbatively Generated Mass}
In SQCD below the scale $\Lambda$ the anomalous dimension $\gamma$ is $ \mathcal{O}\left(1\right)$, and for sufficiently few flavors $\gamma$ is large enough that the meson hits the unitarity bound and becomes a free field. By weakly gauging the flavor symmetry we have introduced two corrections to the mass of the meson, as we can see from Eq.~(\ref{luty}). The first is that the anomalous dimension is corrected by the presence of new gauge interactions giving a term of order $g'^2 m_{UV}^2$. The second is that with a non-zero mass we need to stop the RG evolution at $\mu _*\sim m(\mu_*)$. Dropping the $|F_{\phi}|^2$ term for now, we find that meson mass is determined by its anomalous dimension, which is a function of its Yukawa coupling:
\begin{equation} m_{M'}^2(\mu) = -\frac{1}{3} \left( (F-N) \frac{\lambda_4^2\left(\mu\right)}{8\pi^2}\right)\frac{2F}{b}m_{UV}^2 \end{equation}

 To analyze contributions to the mass associated with the RG running  we separate the contributions to the mass into a $g'=0$ contribution, which is associated with the RG running, and the direct perturbative correction. We saw in section \ref{sec:RG} that near the fixed point, the couplings can be written as their $g'=0$ fixed point values and perturbative corrections, thus
 \begin{equation} m_{M'}^2(\mu) = -\frac{(F-N)}{3} \left( \frac{\lambda^2\left(\mu\right)}{8\pi^2} -a_4\frac{g'^2(\mu)}{8\pi^2}\right)\frac{2F}{b}m_{UV}^2~. \end{equation}
In appendix A we solve the RGEs at $F=3N/2$ in terms of $N$ and $N'$, which gives 
\beq a_4= \frac{\left(3N-2N'\right)\left(N'-1\right)\left(4N^{'2}-N\left(N'+1\right)\right)}{3N^3N'}~.\eeq
Which leads to the mass of the meson $M'$:
\begin{equation} \frac{m_{M'}^2(\mu)}{m_{UV}^2} =  - \frac{N}{2}\frac{\lambda^2\left(\mu\right)}{8\pi^2} +\frac{g'^2(\mu)}{8\pi^2}\left(\frac{\left(3N-2N'\right)\left(N'-1\right)\left(4N^{'2}-N\left(N'+1\right)\right)}{3N^2N'}\right)~.
\label{runningmass}
\end{equation}
Both these terms come from the $D_R$ term in (\ref{luty}); we will see explicitly that the $F_\phi$ term is subleading. 

Next we want to find out when the first term in   (\ref{runningmass}) is parametrically smaller than the second, as a function of $g'$. To do this, we evaluate $m_{M'}$ at an RG scale given by the perturbatively generated mass:
\beq
\mu^2_* =N\,a_4  \frac{g'^2 }{8\pi^2} \,m_{UV}^2~.
\eeq
Now we need to compute approximate solutions to the RGEs\footnote{The RGEs for the couplings of the dual of SQCD are discussed in Appendix A.1}  of $g$ and $\lambda$ in the limit $g'\rightarrow 0$. The RGEs associated with $g$ and $\lambda$ in dual SQCD up to $\mathcal{O}\left((F-N)^{-1}\right)$ corrections are given by:
\beq \frac{dg^2}{d \ln \mu} &=& -\left(\frac{N}{16\pi^2}\right)^2g^4 \left(9\lambda^2-3g^2\right)~,\\
 \frac{ d\lambda^2}{d \ln \mu} &=& \frac{N}{16\pi^2} \lambda^2\left(7\lambda^2-2g^2\right)~.
 \eeq
It is easy to find approximate solutions with the boundary conditons $g\left(\Lambda\right) \rightarrow \infty$ and  $\lambda\left(\Lambda\right)\rightarrow \infty$, which means that, near the matching scale, the dual gauge coupling, and the Yukawa coupling both become very large. The approximate solutions are:
\beq g^2\left(\mu\right) &\approx& \frac{16\pi^2}{N} \sqrt{\frac{7}{6\log\frac{\Lambda}{\mu}}} \\
\lambda^2\left(\mu\right)&\approx&   \frac{2} {7}\,g^2\left(\mu\right) +\frac{3}{14} \frac{N}{16\pi^2} \,g\left(\mu\right)^4
\eeq

Given these approximate solutions we can now compute the parametric dependence of the various mass terms. We find that the RG running term in (\ref{runningmass}) is parametrically given by 
\beq
\ln\left(\frac{\Lambda^{2}}{g'^2m^2_{UV}}\right)^{-1/2}~,
\eeq
while the (so far ignored)   $|F_{\phi}|^2$ term  is suppressed by
 \beq
 \frac{d\gamma}{d\ln \mu} \sim \left(\log\frac{\Lambda}{\mu}\right)^{-3/2}~,
 \eeq
 since we are close to an IR fixed point, and so can be neglected even for $m_{UV}^2\sim | F_{\phi}|^2$.
 
Now we can find a bound on $m_{UV}$ that ensures that the perturbative correction is the dominant one:
 \begin{equation} \lambda^2 \left(\mu_*\right) m_{UV}^2< g'^2  a_4\,m_{UV}^2~.
 \end{equation}
 This gives us a bound on the scale of SUSY breaking relative to the strong scale, $\Lambda$, associated with the original $SU\left(N\right)$. The SUSY breaking mass parameter must obey the following inequality for the perturbative correction is to be the largest contribution:
 \begin{equation} 
 \Lambda^2 \gg m_{UV}^2 >\Lambda^2 \frac{8\pi^2}{ a_4 N g'^2}e^{-64 \pi^4/21(a_{M'} N \,g'^2)^2}~. \end{equation}
This tells us that $m_{UV}$ can be much smaller than the strong scale associated with our original gauge theory. In this range  the leading contribution to the physical mass of the meson is indeed given by:
\begin{equation} m_{M'}^2 \approx N\frac{g'^2}{8 \pi^2}a_4\,m_{UV}^2~. \end{equation}

\subsection{Mass Term and Vacuum}

As we have seen, a small soft SUSY breaking mass for the original quarks generates a mass  for the meson, now we also want to see that this stabilizes the vacuum as well. The potential is given by the sum of two terms, the first comes from a $\ln\left(M'\right)$ term in the superpotential (from $SU(N')$ gaugino condensation) whose coefficient is fixed by the degrees of freedom in the theory, this generates a $\frac{1}{M^{'2}}$ potential. (There will also be quartic terms generated by soft SUSY breaking, but these will be subdominant for small VEVs.) The second term in the potential is the mass term for the meson. Thus we want to find the vacuum given the following form of the potential:
\begin{equation} V\left(M\right) = \frac{N^2}{\left(32\pi^2\right)^2}\frac{\Lambda^{'6}}{M'^2} + \frac{1}{2} \frac{g'^2}{48 \pi^2} a_4\,m_{UV}^2 M'^2~. \end{equation}
The VEV associated with the meson $M'$ is given by  the minimum of this potential:
\begin{equation} \langle M'\rangle = \left(\frac{3N}{32 \pi^2a_4\,g'^2 }\right)^{1/4} \left(\frac{\Lambda^{'3}}{m_{UV}}\right)^{1/2}~. \end{equation}
We see that there is a clear hierarchy of scales within this theory:
\begin{equation} \Lambda'^2 \ll \langle M'\rangle^2 \ll \Lambda^2~. \end{equation}
We also note that the vacuum energy is parametrically small:
\beq
V \sim g' m_{UV} \Lambda^{'3}~.
\eeq

\subsection{Quark Masses}
Using our formalism we can compute the mass of the dual quarks, $q'$. There are two contributions, one is computed directly from the anomalous dimension, as we did for the meson, the second contribution is a mass term generated by the superpotential through the meson VEV:
\beq m_{q'}^2 = -\frac{1}{3}\gamma_{q'}\frac{2F}{b}m_{UV}^2 + \mid \lambda_4\mid^2 \mid \langle M\rangle \mid^2~. \eeq
The anomalous dimension in terms of the Yukawa couplings and the dual gauge couplings the mass becomes:
\beq \gamma_{q'}=\left(\left(F-N'\right)\frac{\lambda_3^2}{8\pi^2} + \frac{\lambda_4^2}{8\pi^2}+\left(N'-1\right)\frac{\lambda_5^2}{8\pi^2} - \frac{g^2}{8\pi^2}\frac{N^2-1}{N} -\frac{g'^2}{8\pi^2}\frac{N'^2-1}{N'}\right)~. \eeq
In terms of the coefficient functions from the RGE this expression is given by:
\beq \gamma_{q'}=\frac{g'^2}{8\pi^2}\left(\left(F-N'\right)a_3 + a_4+\left(N'-1\right)a_5 - a_g\frac{N^2-1}{N}-\frac{N'^2-1}{N'}\right)~. \eeq
Inserting our solutions for $F=\frac{3}{2}N$ we arrive at an expression for the mass of the quarks:
\beq m^2_{q'} = N \frac{g^{'2}}{8\pi^2}\frac{\left(2N+1\right)\left(3N-2N'\right)\left(N'-1\right)\left(N\left(N'+1\right)-4N^{'2}\right)}{3N^4N'} m_{UV}^2 +\mid \lambda_4\mid^2 \langle M\rangle^2 \eeq
In the region in which the meson mass squared is positive, which is approximately $\frac{1}{2}N < N' < \frac{3}{2}N$, the first termis negative, so we need the second term to dominate in order avoid further tachyons. This will place a bound on the scales $\Lambda'$ and the soft SUSY breaking scale $m_{UV}$. The function of $N$ and $N'$ that appears in the first term is bounded below by -2, while the coefficient $a_{4} \sim \mathcal{O}\left(1\right)$. Requiring that the second term dominates we arrive at the bound:
\begin{equation} \Lambda^{'3} > \sqrt{\frac{N}{6\pi^2}}g'm_{UV}^3,\end{equation}
which ensures that the vacuum is stable. Note that the $M'$ VEV , which spontaneously breaks the conformal symmetry also breaks the chiral symmetry of the quarks, and that  the leading contribution to the quark masses is proportional to the dilaton coupling, as we expect from general considerations.

\section{Conclusions}

We have shown that in the context of softly broken and weakly gauged SQCD, at the bottom edge of the conformal window, we can perturbatively generate soft masses for the singlet meson of the weakly coupled IR dual. There is a range of parameters where the leading contribution to the meson mass is the weak gauge coupling itself. Since this meson can play the role of a dilaton in the pure SQCD case when it is set to some arbitrary VEV in the moduli space of vacua, we can recognize that this is a purely 4D example of the Contino-Pomarol-Rattazzi  mechanism: an almost conformally invariant theory where the leading cause of conformal breaking is a perturbation by an almost marginal operator. 
This also provides a concrete example of the dynamics conjectured in ref. \cite{Appelquist:1997fp}, where weakly gauging the flavor symmetry of a non-SUSY CFT leads to spontaneous chiral symmetry breaking. The introduction of soft masses is necessary to  stabilize the vacuum and produce a finite, nonzero VEV for the meson/dilaton in our SUSY example, otherwise the a dilaton could only correspond to a massless flat direction. The meson VEV can be arranged to be larger than the intrinsic scale of the perturbative gauge coupling, and much smaller than the intrinsic scale of the strong interactions, so that all the calculations are self-consistent. While we have concentrated on the special case $F=3N/2$, our arguments are quite general, mainly relying on the weak coupling of the dual theory, so they can be easily extended some distance into the conformal window, i.e. for large $N$ with $F=3N/2+\epsilon N$. It would also be interesting to apply the RG analysis of soft masses to 
phenomenological models of SUSY where composites of strongly coupled gauge theories appear in the low-energy theory, as happens in \cite{Csaki:2012fh} where the top quark is actually a meson of strong dynamics and the stop mass is small due to conformal sequestering, relying crucially on the condition $F=3N/2$. In that case the perturbing weak gauge coupling is actually QCD, and it will be interesting to also consider the effects of the gluino mass, which tends to increase the stop mass.

\section*{Acknowledgements}

We would like to thank Markus A. Luty  and Duccio Pappadopulo for helpful discussions. J.T. is supported in part by the DOE under grant DE-SC-000999.

\appendix
\section{Renormalization Group Equations}

In this section we give the explicit form of the RGEs for the gauge coupling, and Yukawa couplings in the magnetic theory. It is instructive to first look at the case in which $g'=0$ as this will give us a sanity check for our RGEs.

There are two couplings in the dual of SQCD, the gauge coupling associated with the $SU\left(F-N\right)$ gauge group, as well as the Yukawa coupling which couples the dual quarks to the mesons. Rescaling the couplings, $x=\tilde{N}\frac{\lambda^2}{8\pi^2}$ and $y=\tilde{N}\frac{g^2}{8\pi^2}$, where $\tilde{N}=F-N$, we find the RGEs for $x$ and $y$ at leading order:
\begin{equation} \frac{dy}{d \ln \mu} = -\frac{y^2}{\tilde{N}}\left(3\tilde{N}-F+x\frac{F^2}{\tilde{N}}-Fy\left(1-\frac{1}{\tilde{N}^2}\right)\right)\end{equation}
\begin{equation} \frac{dx}{d\ln \mu} = x\left(x\left(1+2\frac{F}{\tilde{N}}\right)-2y\left(1-\frac{1}{\tilde{N}^2}\right)\right) \end{equation}
In the RGE for the gauge coupling we have neglected the denominator in the NSVZ $\beta$-function. As we are looking for weekly coupled fixed points this is not a crucial omission.

We now derive the RGEs associated with our weakly gauged version of SQCD. It is very important, as a sanity check that when we turn off the additional gauge coupling that all the RGEs reduce to their appropriate limit; that is, as $g'\rightarrow 0$ all the Yukawa couplings become equal and should reduce to the above RGE for $x$. All RGEs should be independent of $N'$ in this limit. We similarly rescale the couplings, making the identifications $u=\tilde{N}\frac{g^2}{8\pi^2}$,  $v=\tilde{N}\frac{\lambda_1^2}{8\pi^2}$,  $w=\tilde{N}\frac{\lambda_2^2}{8\pi^2}$,  $x=\tilde{N}\frac{\lambda_3^2}{8\pi^2}$,  $y=\tilde{N}\frac{\lambda_4^2}{8\pi^2}$,  $z=\tilde{N}\frac{\lambda_5^2}{8\pi^2}$ and $\alpha' = \tilde{N}\frac{g'^2}{8\pi^2}$

\begin{equation} \frac{du}{d \ln \mu} = -\frac{u^2}{\tilde{N}}\left(3\tilde{N}-F + v\frac{\left(F-N'\right)^2}{\tilde{N}} +\left(x+w\right)\frac{\left(F-N'\right)N'}{\tilde{N}} +y\frac{N'}{\tilde{N}} + z\frac{N'\left(N'-1\right)}{\tilde{N}}\right)\end{equation}
$$-\frac{u^2}{\tilde{N}} \left(-uF\left(1-\frac{1}{\tilde{N}^2}\right)-\alpha \frac{N'^2-1}{\tilde{N}}\right)$$

\begin{equation} \frac{dv}{d\ln \mu} = v\left(v\left(1+2\frac{F-N'}{\tilde{N}}\right) + \left(x+w\right)\frac{N'}{\tilde{N}} -2u\left(1-\frac{1}{\tilde{N}^2}\right)\right)\end{equation}

\begin{equation} \frac{dw}{d\ln \mu} = w\left(\left(v+x\right)\frac{F-N'}{\tilde{N}} + w\left(1+\frac{N'}{\tilde{N}}\right) +\frac{y}{\tilde{N}} +z\frac{N'-1}{\tilde{N}} -2u\left(1-\frac{1}{\tilde{N}^2}\right)-\alpha \frac{N'^2-1}{N'\tilde{N}}\right)\end{equation}

\begin{equation} \frac{dx}{d\ln \mu} = \frac{dw}{d\ln\mu} \left(w \leftrightarrow x\right) \end{equation}

\begin{equation} \frac{dy}{d \ln \mu} = y\left(y\left(1+\frac{2}{\tilde{N}}\right)+2z\frac{N'-1}{\tilde{N}}+\left(x+w\right)\frac{F-N'}{\tilde{N}}-2u\left(1-\frac{1}{\tilde{N}^2}\right)-2\alpha \frac{N'^2-1}{N'\tilde{N}}\right)\end{equation}

\begin{equation} \frac{dz}{d \ln \mu} = z\left(z\left(1+2\frac{N'-1}{\tilde{N}}\right) + 2\frac{y}{\tilde{N}} + \left(x+w\right)\frac{F-N'}{\tilde{N}} -2u\left(1-\frac{1}{\tilde{N}^2}\right)-2\alpha \frac{2N'^2-1}{N'\tilde{N}}\right)\end{equation}

We also will  list the solutions of the perturbative contributions to the fixed points of the  the RGE for $F=\frac{3}{2}N$. The fixed point solutions are of the form $u = u_* + a_g \alpha'$. For $F=\frac{3}{2}N$ and $g'=0$ the fixed point value is zero, so all solutions are $\mathcal{O}\left(\alpha'\right)$. The $a_i$ each correspond to a fixed point solution, $a_g$ is the $\mathcal{O}\left(\alpha'\right)$ correction to the fixed point value of the dual gauge coupling, and each $a_i$ is for the Yukawa coupling associated with each meson.

\begin{equation} a_g=\frac{4\left(N'-1\right)\left(4N^{'2}-N\left(N'+1\right)\right)}{3N\left(N^2-4\right)}\end{equation}
\begin{equation} a_1=\frac{2\left(N'-1\right)\left(4N^{'2}-N\left(N'+1\right)\right)}{3N^3}\end{equation}
\begin{equation} a_2=a_3=\frac{\left(3N-4N'\right)\left(N'-1\right)\left(N\left(N'+1\right)-4N^{'2}\right)}{6N^3N'}\end{equation}
\begin{equation} a_4=\frac{\left(3N-2N'\right)\left(N'-1\right)\left(4N^{'2}-N\left(N'+1\right)\right)}{3N^3N'}\end{equation}
\begin{equation} a_5=\frac{\left(N^3-4N'\right)\left(8\left(N'-1\right)N^3+2NN'\left(1+\left(6-7N'\right)N'\right)+3N^2\left(5N'^2-1\right)\right)}{3N^4\left(N^2-4\right)N'}\end{equation}

\end{document}